# Simulating the Spread of Influenza Pandemic of 1918-1919 Considering the Effect of the First World War

Teruhiko Yoneyama
Multidisciplinary Science
Rensselaer Polytechnic Institute
Troy, New York, Unites States
yoneyt@rpi.edu

Mukkai S. Krishnamoothy
Computer Science
Rensselaer Polytechnic Institute
Troy, New York, United States
moorthy@cs.rpi.edu

*Abstract*—The Influenza Pandemic of 1918-1919, also called Spanish Flu Pandemic, was one of the severest pandemics in history. It is thought that the First World War much influenced the spread of the pandemic. In this paper, we model the pandemic considering both civil and military traffic. We propose a hybrid model to determine how the pandemic spread through the world. Our approach considers both the SEIR-based model for local areas and the network model for global connection between countries. First, we reproduce the situation in 12 countries. Then, we simulate another scenario: there was no military traffic during the pandemic, to determine the influence of the influenced of the war on the pandemic. By considering the simulation results, we find that the influence of the war varies in countries; in countries which were deeply involved in the war, the infections were much influenced by the war, while in countries which were not much engaged in the war, the infections were not influenced by the war.

I. INTRODUCTION

An influenza pandemic occurred during 1918-1919. It is estimated that about 500 million people (one third of the world's population) were infected and about 50-100 million people died, resulting in the high death rate of 2.5-5% [1][2][3][4]. It is believed that the pandemic originated in the United States, although there are other hypotheses on the origin of the pandemic [5][6][7][8].

One characteristic of 1918-1919 Influenza was that the death rates for the age group 20 to 30 years were remarkable higher than other age groups [4]. Usually the death rate of patients in this age group is low because of their stronger immune system. Figure 1 shows the influenza death rates by age group in 1918 and 1911-1917.

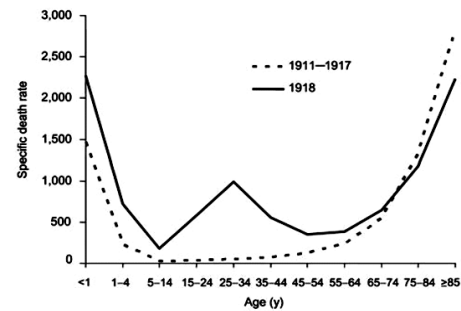

Figure 1: Death Rate of Influenza by Age Group during 1911-1917 (dotted-line) and 1918 (solid-line) [4]

We hypothesize that the spread of the pandemic is based on the traffic pattern. Thus, in this paper, we simulate 12 countries, considering the international traffic in real data. For the international traffic, we consider both the civil traffic and the military traffic since there was the First World War (1914-1918) during the pandemic. We compare the simulation result with the real record on the number of death cases and find important factors which would influence the pandemic. Also, we simulate without military traffic. By comparing the simulation result with the original simulation result, we find how the war influenced the pandemic.

To model the pandemic, we propose a hybrid model which considers both local infection and global infection. For the local infection, we use the SEIR model considering the each country's condition such as domestic population and population density. For the global infection, we use network based model considering the international traffic between countries.

## II. RELATED RESEARCH

Simulations for the spreading of infectious disease have been carried out in the past. There are some differences between our approach and other related research. First, a lot of research on simulating disease spread focuses on a prevention/mitigation strategy by comparing the base simulation and an alternative simulation which considers their proposed strategy (e.g. [9][10][11][12][13][14]). In addition, a lot of existing research simulates with a generated situation which models the real world (e.g. [9][10][13][14][15][16][17]). Our work focuses on the reproduction of the real pandemic using real situation. We model the pandemic, compare the results with real data, and explore the key factors which influenced the spread. Although these critical-factors could provide hints that would help contain the spread of the disease, this paper does not directly propose a prevention strategy.

Earlier research tended to consider the spread of infectious disease from either the local or global point of view (e.g. [11][13][14][16][18]). In addition, much research simulate using one of the equation based (e.g. SIR or SEIR differential equation model), agent based, or network based model (e.g. [16][19][20]). We simulate the pandemic from the global point of view considering local infection in each country. We use a hybrid model which considers both the SEIR based model and network based model using a concept of agent based model.

Third, simulation parameters determine the path of spread. Some earlier research values the basic reproduction number $R_0$ as influential parameter to spread (e.g. [11][21]). We do not determine $R_0$. In our simulation, first, we consider setting the parameters so that the result corresponds with the actual situation in some countries in terms of the number of cases. Then we simulate further experiments using same set of parameters. This is based on the assumption that $R_0$ varies according to country.

## III. MODELING

Previous attempts to model spreading infectious diseases tended to use one of two approaches. Equation-based models like the SEIR model is suitable for a large-scale spreading of diseases. These models use just a few parameters to reproduce the spreading phenomenon. However it is difficult to reflect detailed situation in countries which have different local infection conditions. Second, network or agent-based simulation models can theoretically reflect the detail of individual conditions. However, modeling large-scale global diseases is difficult as too many parameters are needed for simulation. Thus we propose a hybrid model. We make a simple model using a small number of parameters and make it capable of simulating a general pandemic.

We simulate using several countries. When we think of an infection in a country, there are three possibilities for new infection; (1) infection from foreign travelers, (2) infection from returning travelers, and (3) infection from local persons. Figure 2 illustrates this concept. We denote the infection-types (1) and (2) as the global infection and the infection-type (3) as the local infection.

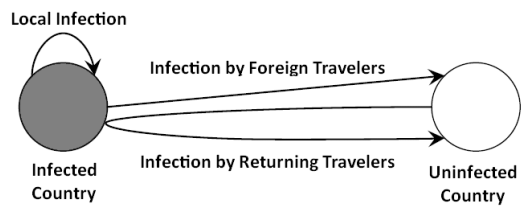

Figure 2: Three Patterns of Infection in a Country

We use the concept of SEIR model which considers four types of agents in each country; Susceptible, Exposed, Infectious, and Removed. Susceptible agents are infected by Infectious agents and become Exposed agents. Exposed agents are in an incubation period. After that period, Exposed agents become Infectious agents. Infectious agents infect Susceptible agents. Infectious agents become Removed agents after the infectious period. Removed agents are never infected again because they are now immune. Figure 3 illustrates this concept of SEIR model.

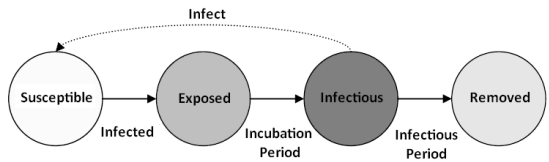

Figure 3: Concept of SEIR

At the beginning of the simulation, the number of Susceptible agents in each country is equal to the population of each country. Then we place an Infectious agent in the origin of the pandemic (i.e. the United States). The local infection spreads around the origin and the global infection also spreads from the origin to other countries through global traffic. When a country has at least one Infectious agent, that country has the potential for local infection. Figure 4 shows this concept.

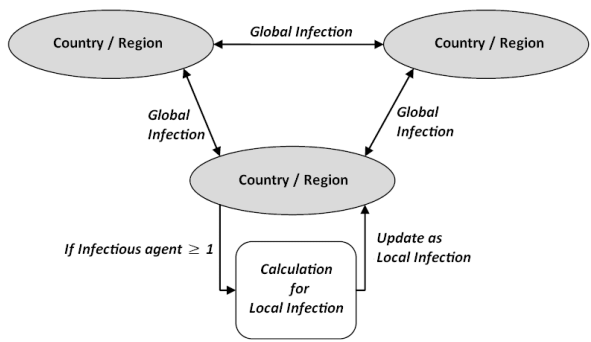

Figure 4: Concept of Simulation Task at One Cycle

The global infection is caused by traffic from infected country. Thus we refer to the number of inbound and outbound traffic. The number of new Exposed agents by the global infection in country $i$ at time $t$, $NEG_i(t)$, is calculated by the expression;

$$NEG_i(t) = I_j(t) \cdot T_{ij} \cdot P_G^*(t) \quad (1)$$

where $I_j(t)$ is the number of Infectious agents of country $j$ at time $t$. $T_{ij}$ is the total amount of both traffic from country $i$ to $j$ and from $j$ to $i$. $P_G^*(t)$ is the global infection probability at time $t$ and is calculated by the expression;

$$P_G^*(t) = P_G - (D_G \cdot t) \quad (2)$$

where $P_G$ is the basic global infection probability between countries. $D_G$ is a "deductor" for the global infection. $t$ is time (simulation cycle). $P_G$ and $D_G$ are constants and are uniformly used for every country. Thus the global infection probability $P_G^*(t)$, decreases along the simulation cycle. We assume that, in the real world, the global infection occurs with high probability in early pandemic due to the lack of awareness of the disease. As the disease spreads, people take preventive measures against the infection and pandemic decreases. We apply this concept in the simulation. The number of Exposed agents in country $i$ at time $t$, $E_i(t)$, is updated by adding $NEG_i(t)$ to $E_i(t)$ at each simulation cycle.

We assume that the local infection probability depends on the population density of a country. Thus if the country is dense, people are more likely to be infected. The basic local infection probability of country $i$, $P_{Li}$ is given by the expression;

$$P_{Li} = Density_i \cdot C_1 + C_2 \quad (3)$$

where $Density_i$ is population density of country $i$, obtained by real data. Thus $Density_i$ differs in country. $C_1$ and $C_2$ are constants and are uniformly used for every country.

We assume that the number of new Exposed cases of a country by the local infection depends on the number of Susceptible agents and the number of Infectious agents at that time. Thus the number of new Exposed agents by the local infection in country $i$ at time $t$, $NEL_i(t)$, is calculated by the expression;

$$NEL_i(t) = S_i(t) \cdot I_i(t) \cdot P_{Li}^*(t) \quad (4)$$

where $S_i(t)$ us the number of Susceptible agents of country $i$ at time $t$. $I_i(t)$ is the number of Infectious agents of country $i$ at time $t$. $P_{Li}^*(t)$ is the local infection probability at time $t$ and is calculated by the expression;

$$P_{Li}^*(t) = P_{Li} - (D_L \cdot t) \quad (5)$$

where $P_{Li}$ is the basic local infection probability of country $i$ which is obtained by equation (3). $D_L$ is a "deductor" for the local infection and is a constant which is used for every country. $t$ is time (simulation cycle). Similar to the global infection, the local infection probability $P_{Li}^*(t)$ decreases as the simulation cycle increases. This reflects people's awareness. The number of Exposed agents in country $i$ at time $t$, $E_i(t)$, is updated by adding $NEL_i(t)$ to $E_i(t)$ at each simulation cycle.

Table 1 lists the parameters in the simulation. There are eight controllable parameters which are denoted as constants in Table 1. These parameters are used for every country uniformly. Other parameters are derived from real data and depend on country.

Table 1: Parameters in Simulation

| Parameter | Description | Attribution | |
|---|---|---|---|
| | | (a) Global or (b) Local | (1) Constant or (2) Depend on Country |
| $P_G$ | Global Infection Probability | (G) | (1) |
| $P_{Li}$ | Local Infection Probability of County $i$ | (L) | (2) |
| $D_G$ | Deductor for Global Infection Probability | (G) | (1) |
| $D_L$ | Deductor for Local Infection Probability | (L) | (1) |
| $C_1$ | Constant for Local Infection Probability | (L) | (1) |
| $C_2$ | Constant for Local Infection Probability | (L) | (1) |
| $Incubation\_Period$ | Incubation Period | (G) and (L) | (1) |
| $Infectious\_Period$ | Infectious Period | (G) and (L) | (1) |
| $Run\_Cycle$ | Run Cycle of Simulation | (G) and (L) | (1) |
| $Density_i$ | Actual Population Density of Country $i$ | (L) | (2) |
| $Population_i$ | Actual Population of Country $i$ | (L) | (2) |
| $T_{ij}$ | Amount of Traffic between Country $i$ and $j$ | (G) | (2) |

IV. SIMULATION AND RESULTS

We select some countries for simulation. At first we select the United States as the origin of the pandemic, and then examine countries strongly related to the United States in terms of amount of trade referring to [26] and select United Kingdom, Canada, Italy, France, and Japan. Next we include India and China since these countries had by far the largest number of deaths compared with other countries. Next, we include Australia and New Zealand since these countries sent their troops to European battlefields and such troops returned to their home countries after the war. We examine the influence of such returning troops upon the spread of the pandemic. Also, we include Russia to examine the effect of the Allied Intervention to Russia for the spread. In addition, we include Switzerland since influenza was reportable disease in Switzerland at that time and the data should be reliable. Thus we simulate these 12 countries.

For the global traffic, we consider both civil traffic and military traffic. Although the war was going on during the pandemic, there was still traffic at the civil level such as trade and travel. Since we don't have exact data on the number of civilian travelers during that period, we refer to the amount of trade between countries, instead. We expect that people travel more when the level of trade is larger between two countries, since the number of people who engage in the shipping is larger. Thus we assume that the situation of trade shows the international relationship at that time. Then we consider the main trading partners of a country and its level of trade to understand the relationship between countries [24][25][26].

In addition to civilian traffic, we consider the military traffic for the international traffic. We focus on three types of military traffic of the First World War which could influence the influenza pandemic of 1918-1919. The most significant factor is the entry of the United States into the war. About 2 million men were sent to Europe during 1918 as American Expeditionary Forces (AEF) [22][23]. A large portion of AEF was sent to France, and some of them were sent to United Kingdom or Italy. In fact, at the beginning of the pandemic in Europe, the infection spread among European troops which associated with U.S. troops [7]. Thus the traffic of AEF much contributed to the spread to Europe. We consider the traffic of AEF referring to [22][23]. Figure 5 shows the number of AEF sailing each month to France and to the United States. Second, some countries besides the United States also sent their troops to European battlefield. Such troops returned to their home countries after the war, during late 1918 through 1919, that corresponds with the pandemic period. Thus it is possible that the returning troops contributed to the spread of the pandemic. We examine the number of returning troops after the war as Table 2 shows. Third, there was the Allied Intervention to Russia from 1918. This dispatch of troops is another major instance of troop-traffic. Thus, it is possible that Russia was infected through such troops from Europe [7]. We examined the number of troops sent to Russia as Table 3 shows. Although the war had worldwide battlefield, the battles outside Europe were almost completed in 1918, except in some parts such as Eastern Africa. Thus, for the military traffic in our simulation, we focus on these three types of troops; AEF, returning troops from Europe, and troops for the Allied Intervention to Russia.

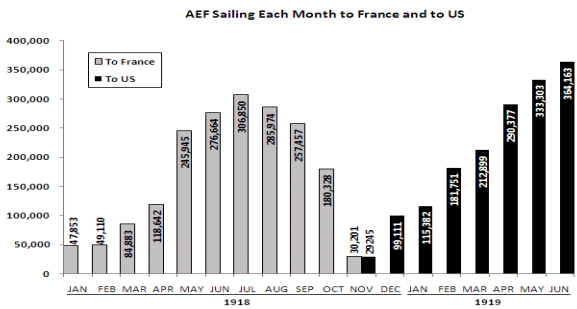

Figure 5: AEF Sailing Each Month to France and to US [7]

Table 2: Estimated Number of Troops Returned from Europe

| Country | Estimated Number of Troops from Europe | Sources |
|---|---|---|
| United States | >1,870,000 from France to US | [22][23][27][28] |
| France | NA | |
| United Kingdom | >1,000,000 from France to UK | [23][29][30][31][32] |
| India | >1,000,000 from France to India | [30][31][33][34][35] |
| Canada | 400,000 from France to Canada | [34][36] |
| Australia | 95,000 from France to England | [34][37] |
| | 195,386 (including dependents) from England to Australia | |
| New Zealand | 90,000 from France to New Zealand | [34][38] |

Table 3: Number of Troops Sent to Russia (as of December 1918)

| | Eastern Russia | | Western Russia | Total |
|---|---|---|---|---|
| | Mrumansk | Archangel | Vladivostok | |
| British and Canadian | 6,832 *1) | 6,293 *1) | 2,000 *3) | 15,125 |
| (Canadian) | | 585? *2) | 4,000 *3) | 4,585 |
| American | NA | 5,302 *1) | 9,014 *4) | 14,316 |
| French | 731 *1) | 1,686 *1) | 500 *1) | 2,917 |
| Italian | 1,251 *1) | NA | NA | 1,251 |
| Japanese | NA | NA | 72,000 *4) | 72,000 |
| Chinese | NA | NA | 2,000 *5) | 2,000 |

*1) [39], *2) [40], *3) [41], *4) [42], *5) [31]

For the local infection, we use the actual population and population density of that time in each country referring to [3][24][25][26][43]. For the global infection, since we have two types of different international traffic, namely civilian and military, we set different global infection probability for each type in our simulation. We set the simulation cycle as 240 to simulate 2 years. Thus we consider 10 cycles in the simulation as one month in the real world. Then we apply the number of troops which travels between countries in the simulation along the time sequence in the real data.

We use the real data on the number of deaths by the pandemic to compare it with the simulation result. Johnson and Mueller collected the real records and comprehensively estimated the mortality rate and death toll in each country [3]. We use this data to compare with our simulation result.

Figure 6 shows the comparison of the number of death cases between simulation result and the real data. Figure 6 (a) shows the comparison in 12 countries and in Figure 6 (b), we eliminate the most two significant countries, India and China, due to the different scale. The simulation result shows that there are a large number of cases in China and India. This tendency corresponds with that of the real world, although there are big differences between simulation result and the real data. In Figure 6 (b), there is substantial difference in Russia. However, in many countries, the number of cases almost corresponds with that of the real data.

We consider the reason for the large differences between the simulation result and the real data in India, China, and Russia. At first, we refer to usual death rate since we guess that usual death rate is reflected by the usual sanitary level in a country. Table 4 shows the vital statistics of deaths per 1,000 persons in 12 countries in 1917. In China and India, the death rate is very high compared with other countries. This may be due to poor sanitary practices in these two countries in the 1910's, which resulted in the large number of deaths.

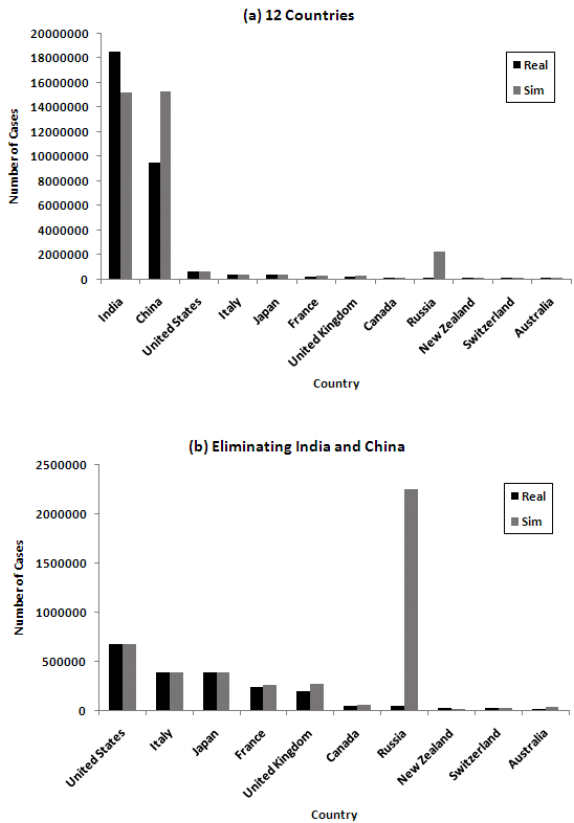

Figure 6: Comparison of Number of Death Cases ((a) In 12 Countries, (b) Eliminating India and China)

Table 4: Vital Statistics of Deaths per 1,000 Persons in 12 Countries in 1917 [24][25][26]

| Country | Deaths per 1,000 |
| --- | --- |
| Australia | 9.7 |
| Canada | 12.7 |
| China *1) | 38 |
| France | 18 |
| India | 32.9 |
| Italy | 26 |
| Japan | 22.2 |
| New Zealand | 9.6 |
| Russia *2) | 25.1 |
| Switzerland | 13.7 |
| United Kingdom | 14.2 |
| United States *3) | 16 |

*1) No data available for 1917-1919. The value is estimated from data in 1949-1969.
*2) Calculated based on data in 1913 and 1926. No exact data for 1917.
*3) Average of Whites and African Americans.

Next, we consider the reason why the real data of India shows larger number of death cases than that the simulation result shows and the real data of China shows smaller number of cases. We refer to the share of age groups in population of a country since the mortality rate of the influenza in 1918-1919 varies in the age group [4]. We focus on the age group of 5-54 since the death rate in this age group was significantly influenced by the influenza compared with the tendency in 1911-1917 as Figure 1 shows. Figure 7 shows the age group of 5-54 in 12 countries. In India, the share of age group of 5-54 is 79.36%. This value is relatively larger percentage among the 12 countries. Thus it might be possible that India had much potential to have a big impact by the pandemic than other countries had, because of the larger share of susceptible age groups. On the other hand, in China, the share of the age groups is 73.55%. This data is in 1953 since there is no data available for China in early 20$^{th}$ century. However, if we assume that the share of each age group in 1953 is similar to that in 1918, it might be possible that China had less potential to have a big impact than India had. Thus this different share of age group brings the difference between the simulation result and the real data in India and China. Since both India and China had large population, such a large population might much increase the influence by the different share of age group. Further we consider that the massive traffic of returning troops to India might have affected the pandemic more than the simulation computed, while China was not significantly involved in the war. In addition, it is possible that the commercial vessels stopped by Indian ports in the trade affected the pandemic since India had many trading partners and was positioned in the major ocean route which connected between Europe and Asia [44]. In the real world, stopping at an intermediate port has possibility to spread the pandemic to the port as well as through trade, but our simulation doesn't consider the intermediate ports.

We consider another exception, Russia. For Russia the number of cases in the simulation result is much larger than that in the number estimated from real data. We are uncertain about the accuracy of the data that was available since Russia had a confusing time during the pandemic due to the First World War, the Civil war and the Allied Intervention, besides the pandemic. If we trust the estimate, one possibility is that Russia had a larger population. Since our model values the population, the result for Russia tends to be large. Another possibility is the share of the susceptible age group of 5-54. It is relatively small among 12 countries as Figure 7 shows. Thus it is possible that Russia had less potential to have a big impact.

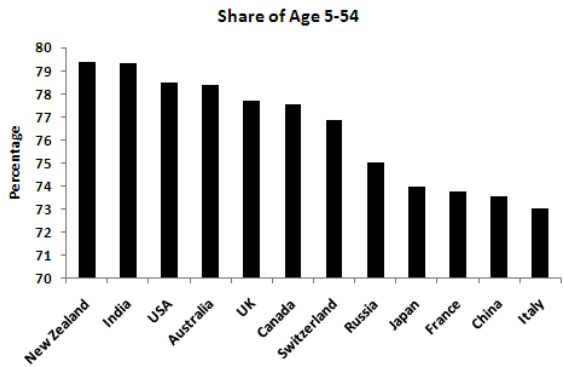

Figure 7: Share of Age Group 5-54 in 12 Countries in 1910's (Created based on [24][25][26], China's data is of 1953)

## V. SIMULATION WITH LESS EFFECT OF THE WWI

Our objective is to determine the influence of the First World War on the pandemic. Now we simulate another scenario. For this scenario, we don't consider all traffic on troops for both the First World War and the Allied Intervention to Russia. We consider only the traffic of trade for the global traffic. For the comparison, we use same parameter values that are used for the previous simulation.

Figure 8 shows the ratio of the number of deaths in the simulation results with no military traffic scenario to that in the simulation result with the original scenario. When a country's ratio is less than 1.0, that country would have smaller number of cases if there was no military traffic. This result shows that Russia, Italy, United Kingdom, and France would have smaller number of cases, if there was no military traffic. These countries received a large number of troops which were sent from the United States or Europe. Thus these countries were infected earlier and had larger number of cases. Russia has the largest difference between scenarios. It is because of the traffic of troops for the Allied Intervention. The difference of the effect by the military traffic varies in countries depending on the amount of trade with the United States, population, and population density of each country.

On the other hand, other countries have no or much less difference between scenarios. For these countries, the war was not a significant factor for the number of death cases. In our simulation, these countries are more likely to be infected through trade rather than through the military traffic. The United States was the origin of the pandemic and Canada was quickly infected by the trade since it is next to the United States. Australia and New Zealand had the returning troops from Europe but these countries had been infected by trade before the troops returned. Although India also received a huge number of returning troops, we consider that the most of the cases in India were due to the local infection because of its large population. China, Japan, and Switzerland were not involved in the European battles. Thus the infection spread only through trade. Therefore in these countries, the timing of infection is not changed even if there is no military traffic, and so there is no or less difference between scenarios.

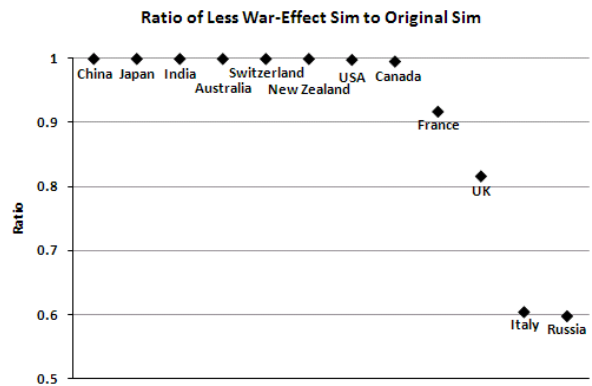

Figure 8: Ratio of the Number of Deaths with Original Simulation Result in 12 Countries

## VI. CONCLUSIONS

In this paper, we modeled the Influenza Pandemic of 1918-1919. For the global infection, we used network based model. We referred to data on trade for civil traffic and considered three types of military traffic: AEF, the returning troops from Europe, and the troops for the Allied Intervention to Russia. For the local infection, we used the SEIR model considering the population and population density of each country. We simulated 12 countries. The result showed similar tendencies to the real data. For some differences between simulation result and the real data, we considered some factors such as usual mortality rate and the share of age group of each country.

We also simulated the scenario without any military traffic to find how the First World War influenced the pandemic. Then we found that the influence of the war upon the pandemic varies in countries. When a country was related with the European battle, the influence by the war was significant. However, when a country was not deeply related with the European battle, the influence on the war was not significant.

REFERENCES

[1] Burnet, F. and Clark, E., *Influenza: A survey of the Last 50 Years in the Loght of Modern Work on the Virus of Epidemic Influenza*, MacMillan, Melbourne, Australia, 1942.

[2] Frost, W.H., *Statistics of Influenza Morbidity*, Public Health Rep. 1920;35:584-97, 1920.

[3] Johnson, N. and Mueller, J., *Updating the Accounts: Global Mortatility of the 1918-1920 Spanish Influenza Pandemic*, Bulletin of the History of Medicine 76, 2002.

[4] Taubenberger, J.K. and Morens, D.M., *1918 Influenza: the Mother of All Pandemics*, Centers for Disease Control and Prevention, Emerging Infectious Diseases Vol.12, No.1, 2006.

[5] Barry, J., *The Site of Origin of the 1918 Influenza Pandemic and Its Public Health Implications*, Journal of Translational Medicine, 2:3, 2009.

[6] Beveridge, W.I., *Influenza: The Last Great Plague, An Unfinished Story of Discovery*, Prodist, New York, USA, 1978.

[7] Crosby, A.W., *America's Forgotten Pandemic: The Influenza of 1918*, Cambridge University Press, Cambridge, UK, 1989.

[8] Hannoun, C., *Documents de la Conference de l'Institut Pasteur: La Grippe Espagnole de 1918*, Maladies Infectieuses 8-069-A-10, 1993.

[9] Carrat, F., et al, *A 'SmallWorldLike' Model for Comparing Interventions Aimed at Preventing and Controlling Influenza Pandemics*, BMC Medicine 2006 4:26, 2006.

[10] Germann, T.C., et al, *Mitigation Strategies for Pandemic Influenza in the United States*, PNAS vol.103 no.15 5935-5940, 2006.

[11] Kelso, J.K., et al, *Simulating Suggests That Rapid Actibation of Social Distancing Can Arrest Epidemic Development Due to a Novel Strain of Influenza*, BMC Public Health 2009 9:117, 2009.

[12] Longini, I.M., et al, *Containing Pandemic Influenza with Antivital Agents*, American Journal of Epidemiology vol.159(7), 2004.

[13] Patel, R., et al, *Finding Optimal Vaccination Strategies for Pandemic Influenza Using Genetic Algorithms*, Journal of Theoretical Biology 234 201-212, 2004.

[14] Weycker, D., et al, *Population Wide Benefits of Routine Vaccination of Children against Influenza*, Vaccine 23 12841293, 2004.

[15] PastorSatorras, R. and Vespignani, A., *Epidemic Dynamics in Finite Size ScaleFree Networks*, Physical Review E vol.65 035108, 2002.

[16] Glass, R.J., et al, *Targeted Social Distancing Design for Pandemic Influenza*, Centers for Disease Control and Prenention, Emerging Infectious Diseases vol.12 no.11 2006.

[17] Eubank, S., *Scalable, Efficient Epidemiological Simulation*, Proceedings of the 2002 ACM symposium on Applied Computing, 2002.

[18] Jenvald, J., et al, *Simulation as Decision Support in pandemic Influenza Preparedness and Response*, Proceedings ISCRAM2007, 2007.

[19] Chowell, G., et al, *Comparative Estimation of the Reproduction Number for Pandemic Influenza from Daily Case Notification Data*, Journal of the Royal Society Interface 2007 4 155166, 2007.

[20] Deguchi, H., et al, *Anti Pandemic Simulation by SOARS*, SICE-ICASE International Joint Conference 2006, pp.18-21, 2006.

[21] Ferguson, N.M., et al, *Strategies for Mitigating an Influenza Pandemic*, Nature vol.442 448-452, 2006.

[22] Banks, A., *A Military Atlas of the First Would War*, Pen and Sword Books, South Yorkshire, UK, 2002.

[23] Ayres, L.P., *The War with Germany: A Statistical Summary*, Washington Government Pritinting Office, Washington D.C., USA, 1919.

[24] Mitchell, B.R., *International historical statistics: Africa, Asia, & Oceania, 1750-1993*, Macmillan Reference Ltd, London, UJ, Stockton Press, New York, USA, 1998.

[25] Mitchell, B.R., *International historical statistics: Europe, 1750-1993*, Macmillan Reference Ltd, London, UJ, Stockton Press, New York, USA, 1998.

[26] Mitchell, B.R., *International historical statistics:The Americas, 1750-1993*, Macmillan Reference Ltd, London, UJ, Stockton Press, New York, USA, 1998.

[27] Leland, A. and Oboroceanu, M-J, *American War and Military Operations Casualities: Lists and Statistics, CRS Report for Congress*, Congressional Research Service, USA, 2009, http://www.fas.org/sgp/crs/natsec/RL32492.pdf

[28] Western Front Association, *Morbilization and Casuality Figures*, http://www.westernfrontassociation.com/great-war-on-land/other-wa-theatres/293-mob-cas-figure.html

[29] Martin, G., *Atlas of World War I*, Oxford University Press, Oxford, UK, 1994.

[30] Commonwealth War Graves Commission, *Annual Report 2007-2008 Online*, http://tinyurl.com/23jua7

[31] Tucker, S.C. and Roberts, P.M., *The encyclopedia of World War I: A Political, Social and Military History*, ABCCLIO, California, USA, 2005.

[32] Army Council, *General Annual Report of the British Army 1912-1919*, Parliamentary Paper 1921 XX Cmd.1193, UK, 1921.

[33] Chappell, M., *The British Army in World War I: The Western Front 1914-16*, Osprey Publishing, Oxford, UK, 2003.

[34] Bean, C., *Official History of Australia in the War of 1914-1918*, http://www.awm.gov.au/histories/first_world_war, Australian War Memorial, Australia, 1920-1942.

[35] Urlanis, B., *Wars and Population*, University Press of the Pacific, Hawaii, USA, 2003.

[36] Veterans Affairs Canada, *The Last Hundred Days*, http://www.vac-acc.gc.ca/remembers/sub.cfm?source=feature/hundreddays

[37] Australian War Memorial, http://www.awm.gov.au/atwar/index/ww1.asp

[38] New Zealand History Online, *New Zealand and the First World War*, http://www.nzhistory.net.nz/war/first-world-war-overview/introduction

[39] Swettenham, J., *Allied Intervention in Russia 1918-1919: And the Port Played by Canada*, Geroge Allen & Unwin Ltd, London, UK, 1967.

[40] National Defence Headquarters of Canada, *Report No.82 Hostorical Section (G.S.) Army Headquarters: Operations in Northern Russia 1918-1919*, Directorate of History, National Defense Headquarters, Ottawa, Canada, 1986.

[41] Smith, G., *Canada and the Siberian Intervention*, 1918-1919, The American Historical Review Vol.64 No.4 (Jul., 1959) pp.866-877, American Historical Association, 1959.

[42] Untergerger, B.M., *American Intervention in the Russian Civil War*, Raytheon Education Company, Massachusetts, USA, 1969.

[43] Press Publishing Co., *The World Almanac ad Encyclopedia 1915*, Press Publishing Co., New York, USA, 1914.

[44] Berglund, A., *Ocean Transportation*, Longmans Green and Co., New York, USA, 1931.